\documentclass[pra,prl,onecolumn,superscriptaddress,floatfix,showpacs,10pt]{revtex4}
\usepackage{amsfonts}
\usepackage{amsmath}
\usepackage{amssymb}
\usepackage{graphicx}

\begin{document}

\title{Entropy is Conserved in Hawking Radiation as Tunneling:
a Revisit of the Black Hole Information Loss Paradox}
\author{Baocheng Zhang}
\affiliation{State Key Laboratory of Magnetic Resonances and Atomic and Molecular
Physics, Wuhan Institute of Physics and Mathematics, Chinese Academy of
Sciences, Wuhan 430071, China}
\affiliation{Graduate University of Chinese Academy of Sciences, Beijing 100049, China}
\author{Qing-yu Cai}
\email{qycai@wipm.ac.cn}
\affiliation{State Key Laboratory of Magnetic Resonances and Atomic and Molecular
Physics, Wuhan Institute of Physics and Mathematics, Chinese Academy of
Sciences, Wuhan 430071, China}
\author{Ming-sheng Zhan}
\affiliation{State Key Laboratory of Magnetic Resonances and Atomic and Molecular
Physics, Wuhan Institute of Physics and Mathematics, Chinese Academy of
Sciences, Wuhan 430071, China}
\affiliation{Center for Cold Atom Physics, Chinese Academy of Sciences, Wuhan 430071,
China }
\author{Li You}
\affiliation{Department of Physics, Tsinghua University, Beijing 100084, People's
Republic of China}

\begin{abstract}
We revisit in detail the paradox of black hole information loss due
to Hawking radiation as tunneling. We compute the amount of
information encoded in correlations among Hawking radiations for a
variety of black holes, including the Schwarzchild black hole, the
Reissner-Nordstr\"{o}m black hole, the Kerr black hole, and the
Kerr-Newman black hole. The special case of tunneling through a
quantum horizon is also considered. Within a phenomenological
treatment based on the accepted emission probability spectrum from a
black hole, we find that information is leaked out hidden in the
correlations of Hawking radiation. The recovery of this previously
unaccounted for information helps to conserve the total entropy of a
system composed of a black hole plus its radiations. We thus
conclude, irrespective of the microscopic picture for black hole
collapsing, the associated radiation process: Hawking radiation as
tunneling, is consistent with unitarity as required by quantum
mechanics.
\end{abstract}

\pacs{04.70.Dy, 04.60.-m, 03.67.-a}

\maketitle

%\email{qycai@wipm.ac.cn}

\section{\textbf{Introduction}}

Since Hawking radiation was first discovered \cite{swh74,swh75}, its
inconsistency with quantum theory has been widely debated. Within the
scenario developed along the original work of Hawking \cite{swh75,swh76},
irrespective of the initial state of the composing matter, it always evolves
into a thermal state after collapsing into a black hole. Such a picture
leads to a paradoxical claim of black hole information loss or the violation
of entropy conservation, as different initial states: pure or mixed, evolve
into the same final thermal state. This directly violates the principle of
unitarity for quantum dynamics of an isolated system and brings a serious
challenge to the foundations of modern physics. The lack of a resolution on
this fundamental problem concerning thermodynamics, relativity, quantum
mechanics, and cosmology, has attracted considerable attention. This problem
is now generally known as \textquotedblleft the paradox of black hole
information loss\textquotedblright. In the past few decades, several methods
\cite{swh76,acn87,kw89,jp92,hm04,swh05} have been suggested for resolving
this paradox; none has been successful. In fact, each failed attempt for a
resolution seems to have made the existence of this paradox more serious and
attracted more interest, especially after the possibility that information
about infallen matter may hide inside the correlations between the Hawking
radiation and the internal states of a black hole was ruled out. It seems
either unitarity or Hawking's semiclassical treatment of radiation must
break down \cite{bp07}.

In most previous resolutions to the black hole information loss paradox \cite%
{swh76,acn87,kw89,jp92,hm04,swh05,bp07}, the radiation from a black hole is
considered purely thermal because the background geometry is fixed and
energy conservation is not enforced during the radiation process. On the
other hand, some have correctly argued that information could come out if
the emitted radiations were not exactly thermal but instead the radiation
spectrum contains a subtle non-thermal correction \cite{jdb93}. The
semiclassical derivation of Hawking is based on Bogoliubov transformation
which always gives pure thermal radiations. Although Hawking later explained
the particles of radiation as stemming from vacuum fluctuations tunneling
through the horizon of a black hole together with Hartle \cite{hh76}, the
original semiclassical treatment did not have any direct connection with
tunneling. It is thus important to review the picture of Hawking radiation
as tunneling, whereby a pair of particles is spontaneously generated inside
the horizon. The positive energy particle tunnels out to the infinity while
the negative energy one remains in the black hole. Alternatively, the
positive and negative energy pair is created outside the horizon, and the
negative energy particle tunnels into the black hole because its orbit
exists only inside the horizon, while the positive energy one remains
outside and emerges at infinity.

Recently, Parikh and Wilczek \cite{pw00} developed the method of Hawking
radiation as tunneling due to Kraus and Wilczek \cite{kw94,kw95}. The flux
of particles can be computed directly within their tunneling picture.
Extensions to many other situations have since been made \cite%
{jz05,jc06,jz06,amv05,nm08,aps,aas}, that established a firm basis
for the physical explanation of Hawking radiation as tunneling. In
their current version \cite{pw00,mkp04,mkp042}, energy conservation
is directly enforced and is observed to play an important role. The
actual calculation benefits from a coordinate system, the
Painlev\'{e} coordinate, which unlike the Schwarzchild coordinate,
is regular at the horizon, and is considered to be a particularly
suitable and convenient choice. The tunneling process involves
particles that can be supplied by considering the geometrical limit
because of the infinite blueshift of the outgoing wave-packets near
the horizon. The barrier is created by the outgoing particle itself,
which is ensured by energy conservation. Finally, the radial null
geodesic motion is considered, and the WKB approximation is used to
obtain a tunneling probability $\Gamma \sim \exp \left[
-2\mathrm{Im}(I)\right] $ related to the imaginary part of the
action $I$ of the black hole. This tunneling
probability is related to the change of the entropy of a black hole \cite%
{pw00}. It is clearly non-thermal. As was pointed out by us in an earlier
communication \cite{zcz09}, this non-thermal feature implies the existence
of information-carrying correlations among sequentially emitted Hawking
radiations.

In a recent communication \cite{zcz09}, we visited the topic of the black
hole information loss paradox starting with the non-thermal spectrum
obtained by Parikh and Wilczek \cite{pw00}. For a typical black hole, a
Schwarzchild black hole, we discover the existence of correlation among
Hawking radiations. Upon carefully evaluating the amount of information
carried away in this correlation, we find there is simply no loss of
information and that the entropy is conserved for the total system of the
black hole plus its radiations. The information coded into the correlations
and carried away by the Hawking radiation is found to balance exactly the
amount claimed lost previously \cite{zcz09}. Our result thus shows that
tunneling through the horizon is an entropy-conserving process when the
entropy associated with the correlations of emitted particles is included.
This conclusion is consistent with the principle of unitarity for quantum
mechanics concerning an isolated system.

The purpose of this article is to revisit the paradox of black hole
information loss, along the ideas originally developed by us in the earlier
communication \cite{zcz09}. In addition to provide more details for an
indepth understanding and to support our original claim, we also extend our
previous analysis from the Schwarzschild black holes to the Reissner-Nordstr%
\"{o}m black holes, the Kerr black holes, and the Kerr-Newman black
holes. For charged black holes, both tunnelings due to massless
neutral particles and massive charged particles will be considered.
We find: (a) sequentially tunneled particles are correlated with
each other, and (b) the entropy of the total system composed of a
black hole and its radiations is conserved provided correlations
between Hawking radiations are included. For massless neutral
particles, due to charge conservation the black hole will evolve
into the extreme case where the temperature is zero and the
radiation vanishes. The total entropy remains conserved during the
evolution. Before the extreme case arrives, all our analysis remain
appropriate and the correlation is found to be capable of taking the
information out from a charged black hole as we suggested. In the
extreme case, the total information remains conserved, although some
may have leaked out while others stay in the stable extreme black
hole. In order to avoid the extreme case we analyze the tunneling of
massive charged particles from the Reissner-Nordstr\"{o}m black
holes and we verify that the entropy conservation still holds due to
energy and charge conservations. For rotating black holes and
charged rotating black holes, the same conclusion is reached:
information is carried out by correlations among the non-thermal
radiations; the total entropy is conserved. Thus the Hawking
radiation as tunneling in the above cases must also be unitary.
Additionally, we discuss Hawking radiation as tunneling through a
quantum horizon where a black hole may evolve into a remnant and not
evaporate at all. This case does not conflict with our claim of
entropy conservation either, as information is still carried away by
the correlations between outgoing particles.

Summarizing our work reported in this paper, for an extensive list
of black holes, we find the inclusion of the correlations hidden
inside non-thermal Hawking radiations presents a resolution to the
paradox of black hole information loss. We find the total entropy is
always conserved, significantly extending our earlier claim
\cite{zcz09}. We thus have provided a self-consistent and reasonable
resolution to the paradox of black hole information loss. At the
heart of our resolution is the existence of correlations among
emissions of Hawking radiation. Independent of the microscopic
picture of how a black hole actually collapses and what its initial
and final states are, a non-thermal emission spectrum implies the
existence of correlation. As we show below in detail, for various
types of black holes, after carefully counting the total
correlation, we find the total entropy is conserved, which implies
that no information loss occurs in the process of Hawking radiation
as tunneling.

Our paper is organized as follows. In section II we first review our
method of resolving the paradox for a Schwarzschild black hole in
the picture of Hawking radiation as tunneling \cite{zcz09}.
Subsequently, this method is applied to the analysis of radiation as
tunneling for Reissner-Nordstr\"{o}m black holes, Kerr black holes,
and Kerr-Newman black holes. In the third section, we discuss
information loss paradox for Hawking radiation as tunneling through
a quantum horizon. Section IV ends with conclusions and some
remarks. To simplify the expressions and calculations, we take the
convenient units of $k=\hbar =c=G=1$.

\section{Hawking radiation as tunneling through a classical horizon}

We start with a brief review of the method and ideas we introduced earlier
for resolving the paradox of black hole information loss \cite{zcz09}. As
was pointed out before, we find that correlations exist among Hawking
radiations from a Schwarzchild black hole if its radiation spectrum is
non-thermal as required by energy conservation. A careful counting of the
total correlations is shown by us to balance exactly the information
previously considered lost. After the review, we will perform analogous
analysis for other situations such as Reissner-Nordstr\"{o}m black holes,
Kerr black holes, and Kerr-Newman black holes.

\subsection{Schwarzchild black hole}

To properly describe any phenomena involves the crossing of horizon, it is
helpful to change from the Schwarzchild coordinates into Painlev\'{e}
coordinates, which are not singular at the horizon. Finding the radial null
geodesic and computing the imaginary part of the action for the process of
s-wave emission across the horizon, the tunneling probability is found to be
\cite{pw00}
\begin{equation}
\Gamma \sim \exp \left[ -8\pi E\left( M-\frac{E}{2}\right) \right] =\exp
\left( \Delta S\right) ,  \label{tb}
\end{equation}%
where the second equal sign expresses this result in terms of the change of
the Bekenstein-Hawking entropy \cite{swh74,swh75,jdb730} for the
Schwarzchild black hole $S_{\mathrm{BH}}=A/4=4\pi M^2$, where $A=4\pi (2M)^2$
is the surface area of a Schwarzchild black hole with mass $M$ and radius $%
2M $. It is important to note that this spectrum is non-thermal, different
from the thermal case of a simple exponential $\Gamma (E)=\exp \left( -8\pi
EM\right) $. A non-thermal spectrum implies that individual emissions are
correlated because the emission probability for two simultaneous emissions
is not the same as the product probabilities for two independent emissions.
This point has been outlined in detail in an earlier communication \cite%
{zcz09}, where we show that such correlations can encode information, thus
leading to information being carried away by Hawking radiations. When the
amount of information carried away by correlation is included, the total
entropy of the system composed of the black hole and Hawking radiation is
conserved. In other words the non-thermal spectrum suggests unitarity and no
information loss in black holes.

In statistical theory \cite{gs92}, if the probability for two events arising
simultaneously is identically equal to the product probability of each event
arising independently, these two events are independent and there exists no
correlation between them. Otherwise, they are correlated. Accepting the
spectrum of Eq. (\ref{tb}), the probability for the first emission at an
energy $E_1$ becomes
\begin{eqnarray}
\Gamma (E_{1}) &=&\exp \left[ -8\pi E_{1}\left( M-\frac{E_{1}}{2}\right) %
\right] .  \label{1tp}
\end{eqnarray}%
After this first emission, the mass of the black hole is reduced to $M-E_1$
due to energy conservation. The conditional probability for a second
emission at an energy $E_2$ is therefore given by
\begin{eqnarray}
\Gamma (E_{2}|E_{1})=\exp \left[ -8\pi E_{2}\left( M-E_{1}-\frac{E_{2}}{2}%
\right) \right].  \label{cp}
\end{eqnarray}
The tunneling probability for two emissions with energies $E_{1}$ and $E_{2}$%
, respectively, can be computed accordingly as
\begin{eqnarray}
\Gamma (E_{1},E_{2})=\Gamma (E_{1})\Gamma (E_{2}|E_{1})&=&\exp \left[ -8\pi
(E_{1}+E_{2})\left( M-\frac{E_{1}+E_{2}}{2}\right) \right].  \label{atp}
\end{eqnarray}%
Interestingly, we find
\begin{eqnarray}
\Gamma (E_{1},E_{2})=\Gamma (E_{1}+E_{2}),  \label{dbj}
\end{eqnarray}
or the tunneling probability of a particle with an energy
$E_{1}+E_{2}$ is the same as the probability for two emissions of
energies at $E_{1}$ and $E_{2}$.

To prove the existence of correlation between the two emissions and
to properly quantify the amount of correlation, we need to find the
independent probability for each emission. Using the theory of
probability, we find the marginal probability $\Gamma_1 (E_{1})$ for
the emission at energy $E_1$ is identically the same as $\Gamma
(E_{1})$ \cite{nt1}. For the emission at energy $E_2$, we find the
marginal probability $\Gamma_2(E_{2})$ again takes the same
functional form of $\Gamma(E_{2})$. We find that
\begin{equation}
\ln \Gamma (E_{1}+E_{2})-\ln \left[\Gamma (E_{1})\ \Gamma (E_{2})\right]
=8\pi E_{1}E_{2}\neq 0,  \label{co}
\end{equation}%
which shows that the two tunnelings are not statistically
independent, and there indeed exist correlations between Hawking
radiations. As will become clear later, the existence of this
correlation is central to the resolution we provide for the black
hole information loss paradox.

A fundamental assumption in statistical mechanics concerns the equal
probability distribution for every micro-state, which forms the basis of the
micro-canonical ensemble approach. Given the quantum tunneling of an emitted
particle with an energy $E$, or a Hawking radiation from a black hole, with
the probability of Eq. (\ref{tb}), when the black hole is exhausted, we can
find the entropy of the total system by counting the numbers of its
microstates. For example, one of the microstates is $\left(
E_{1},E_{2},\cdots ,E_{n}\right) $ and $\sum_{i}E_{i}=M$. Within such a
description, the order of $E_{i}$ cannot be changed, the distribution of
each $E_{i}$ is consistent with the tunneling probabilities discussed in the
main text. The probability for the specific microstate $\left(
E_{1},E_{2},\cdots ,E_{n}\right) $ to occur is given simply by
\begin{eqnarray}
P_{( E_{1},E_{2},\cdots,E_{n})} &=& \Gamma ( E_{1},E_{2},\cdots,E_{n})
\notag \\
&=& \Gamma (E_{1})\times \Gamma (E_2|E_{1})\times \cdots \times \Gamma
(E_{n}|E_{1},E_{2},\cdots E_{n-1}).
\end{eqnarray}

Following the steps involved in arriving at Eq. (\ref{dbj}), it is easy to
show
\begin{eqnarray}
\Gamma (E_{1},E_{2},E_3) &=& \Gamma (E_{1}+E_{2},E_3)=\Gamma
(E_{1}+E_{2}+E_3),
\end{eqnarray}
and analogous identities for all subsequent emissions. Finally we obtain
\begin{eqnarray}
P_{( E_{1},E_{2},\cdots,E_{n})} &=& \Gamma ( E_{1},E_{2},\cdots,E_{n})
\notag \\
&=& \Gamma ( \sum_{j=1}^N E_{j})=\Gamma (M)=\exp (-4\pi M^{2})=\exp (-S_{%
\mathrm{BH}}).
\end{eqnarray}
The total number of microstates is therefore given by
\begin{eqnarray}
\Omega =\frac{1}{P_{( E_{1},E_{2},\cdots,E_{n})}}=\exp (S_{\mathrm{BH}}).
\end{eqnarray}

According to the Boltzmann's definition, the entropy of a system is given by
$S=\ln \Omega =S_{\mathrm{BH}}$, where the Boltzmann's constant is taken as
unity for simplicity. Thus we show after a black hole is exhausted due to
Hawking radiation, the entropy carried away in the emitted particles
(Hawking radiations) is precisely equal to the entropy $S_{\mathrm{BH}}$ in
the original black hole \cite{swh74,swh75,jdb730}.

This result is in direct contradiction with the black hole information loss
paradox. A lot of previous investigations \cite{swh76,rmw76,dnp76,wgu76}
support the claim that information can be lost in a black hole. Thus the
total entropy of a black hole increases during Hawking radiation \cite%
{whz82,dnp83}. The analysis we provide here, however, shows otherwise. Based
on a straightforward calculation using statistical theory, we find the total
entropy is conserved. This shows the time evolution of a black hole is
unitary. In particular, the Hawking radiation remains governed by
conservations laws we are accustomed to.

Two significant points distinguish our investigation from most existing
theories. First, we start with the assumption of the non-thermal spectrum
for Hawking radiation as derived by Parikh and Wilczek \cite{pw00}. Second,
we discover the existence of information-carrying correlations among
different Hawking emissions assuming the nonthermal spectrum \cite{zcz09}.
The physics of both points lie at the fundamental law of energy
conservation. In contrast to the earlier thermal spectrum of Hawking, Parikh
and Wilczek enforced energy conservation when they computed the emission
spectrum based on Hawking radiation as tunneling. The arising of
correlations between different emissions then becomes easy to understand as
we have shown earlier: after the emission of a Hawking radiation, the mass
of the remaining black hole decreases, which subsequently affects the next
emission, thus establishes correlations between different emissions.
Assuming the particular form of the non-thermal spectrum by Parikh and
Wilczek, the correlations among different emissions of Hawking radiation can
be thoroughly studied.

We next provide a careful quantification for the amount of the correlation
between different Hawking emissions in terms of the amount of information it
can encode. In addition to establishing an alternative proof, the discussion
in the next few paragraphs provides an insightful understanding of our claim
that Hawking radiation is a unitary process and the entropy for a black hole
plus its Hawking radiation is conserved. More vividly, our analysis below
shows that Hawking radiations can be viewed as messengers. In this way,
information is leaked out through correlated tunneling processes that will
be shown clearly when we compare the amount of correlation with the mutual
information between the two emissions.

Because of the existence of correlation, we shall be very careful
when considering emissions of particles with energies $E_{1}$ and
$E_{2}$, one after another, because $\Gamma (E_{2}|E_{1})$ is the
conditional probability for an emission at energy $E_{2}$ given the
occurrence of an emission with energy $E_{1}$. More generally we use
$E_{i}$ to denote the energy for the $i$th emission. Given the total
energy for all previous emissions $\sum E_i=E_1+E_2+\cdots+E_{f-1}$,
the tunneling probability for a next emission of energy $E_{f}$
becomes $\Gamma (E_{f}|E_1,E_2,\cdots,E_{f-1})=\exp \left[
-8\pi E_{f}\left( M-\sum E_{i}-\frac{E_{f}}{2}\right) \right] $. From $%
\Gamma (E_{f}|E_1,E_2,\cdots,E_{f-1})$, we can compute the entropy taken
away by a tunneling particle with energy $E_{f}$ after the black hole has
emitted a total energy $\sum E_{i}$, which is given by
\begin{equation}
S\left( E_{f}\left|E_1,E_2,\cdots,E_{f-1}\right)\right. =-\ln \Gamma
\left(E_{f}\left|E_1,E_2,\cdots,E_{f-1}\right)\right..  \label{te}
\end{equation}

In quantum information theory, $S\left( E_{f}|E_{1},E_{2},\cdots
,E_{f-1}\right) $ denotes conditional entropy and is used to
quantify the remaining entropy of an emission with energy $E_{f}$
given that information for all previously emitted particles with a
total energy $\sum E_{i}$ are known. In quantitative terms, we find
$S(E_{f}|E_{1},E_{2},\cdots ,E_{f-1})$ is equal to the decrease of
entropy for a black hole with a mass $M-\sum E_{i}$ upon an emission
of $E_{f}$. This also agrees with the general second law of black
hole thermodynamics \cite{jdb73,swh762}. The tunneling particles
must carry entropy with themselves because the total entropy of a
black hole and its radiations can never decrease. In what follows we
will show that by using entropy we can measure the amount of
information hidden in the correlation (\ref{co}).

The mutual information \cite{nc00} between two subsystems $A$ and $B$ in a
composite bi-partite system is defined as%
\begin{equation}
S(A:B)\equiv S(A)+S(B)-S(A,B)=S(A)-S(A|B),
\end{equation}%
where $S(A|B)$ is nothing but the conditional entropy. This mutual
information can be used to measure the total amount of correlations between
any bi-partite systems. When applied to the emission of two particles with
energies $E_{1}$ and $E_{2}$, their mutual information becomes
\begin{equation}
S(E_{2}:E_{1})\equiv S(E_{2})-S(E_{2}|E_{1})=-\ln \Gamma (E_{2})+\ln \Gamma
(E_{2}|E_{1}).  \label{tmi}
\end{equation}

For a classical horizon, using Eqs. (\ref{1tp}) and (\ref{cp}), we find
previously $S(E_{2}:E_{1})=8\pi E_{1}E_{2}$. This shows that the correlation
between emissions of Hawking radiation can carry information, \emph{i.e.},
the above Eq. (\ref{tmi}) affirms that the amount of correlation quantity in
Eq. (\ref{co}) is precisely equal to the mutual information between
emissions for a classical horizon. Additionally, this justifies the
reexamination of the entropy (\ref{te}) by quantum information theory.

We now compute the entropy of all tunneled particles. The entropy for the
first tunneled particle, with energy $E_{1}$ from a black hole of mass $M$,
is given by
\begin{equation}
S(E_{1})=-\ln \Gamma (E_{1})=8\pi E_{1}\left( M-\frac{E_{1}}{2}\right).
\label{tp1}
\end{equation}%
The entropy for the second tunneling particle with energy $E_{2}$ after the
emission of a particle with energy $E_{1}$ becomes
\begin{equation}
S(E_{2}|E_{1})=-\ln \Gamma (E_{2}|E_{1})=8\pi E_{2}\left( M-E_{1}-\frac{E_{2}%
}{2}\right) ,  \label{ctp}
\end{equation}%
analogous to the formula (\ref{tp1}), except for the reduced mass of the
black hole to $M-E_{1}$ due to the first emission from energy conservation.
The total entropy from the two emitted particles $E_{1}$ and $E_{2}$ is
\begin{equation}
S(E_{1},E_{2})=S(E_{1})+S(E_{2}|E_{1}),
\end{equation}%
\textit{i.e.}, rightfully including the contribution from their correlation.
This result can be repeated. After the tunneling of particles with energies $%
E_{1}$ and $E_{2}$, the mass of the black hole becomes $M-E_{1}-E_{2}$, and
it proceeds to emit a third particle with energy $E_{3}$ due to tunneling.
The corresponding tunneling entropy is $S(E_{3}|E_{1},E_{2})=-\ln
(E_{3}|E_{1},E_{2})$, which gives the total entropy for the three emissions,
respectively, at energies $E_{1}$, $E_{2}$, and $E_{3}$,
\begin{equation}
S(E_{1},E_{2},E_{3})=S(E_{1})+S(E_{2}|E_{1})+S(E_{3}|E_{1},E_{2}).
\end{equation}
For all emissions which eventually exhausts the black hole, we find
\begin{equation}
S(E_{1},E_{2},...,E_{n})=\sum\limits_{i=1}^{n}S(E_{i}|E_{1},E_{2},\cdots
,E_{i-1}),  \label{bhe}
\end{equation}%
where $M=\sum_{i=1}^{n}E_{i}$ is the initial energy of the black hole due to
energy conservation. The generalized term $S(E_{1},E_{2},...,E_{n})$ denotes
the joint entropy of all emitted radiations and $S(E_{i}|E_{1},E_{2},\cdots
,E_{i-1})$ is the respective conditional entropy for the $i$th emission with
energy $E_{i}$ after the emissions of a total of $i-1$ particles. We recall
that Eq. (\ref{bhe}) satisfies the chain rule for conditional entropies in
information theory (Please see Ref. \cite{nc00}, chapter 11). When the black
hole is exhausted, all of its entropy is carried away by Hawking radiations.
By a detailed calculation from Eq. (\ref{bhe}), we previously show that the
total entropy of all Hawking radiations equals to $%
S(E_{1},E_{2},...,E_{n})=4\pi M^{2}$, which is exactly the same as the
Bekenstein-Hawking entropy of a black hole. This equation shows that the
entropy of a black hole is indeed taken out by Hawking radiations, and the
total entropy of all emitted radiations and the black hole is unchanged
during the black hole radiation process. According to quantum mechanics,
only unitary processes conserve the entropy for a closed system. Thus we
conclude that irrespective of the microscopic picture for a black hole, the
fact that the total entropy remains conserved during Hawking radiation
implies that the black hole radiation as tunneling is unitary in principle.
This provides a self-consistent and reasonable resolution to the long
standing paradox of black hole information loss.

\subsection{Reissner-Nordstr\"{o}m black hole}

The method of null geodesic can also be applied to treat Hawking radiation
from a charged black hole. The emission due to tunneling of non-charged
(neutral) particles was considered in Ref. \cite{pw00}. The counterpart to
the Painlev\'{e} coordinate for the charged Reissner-Nordstr\"{o}m
coordinate is
\begin{equation}
ds^{2}=-\left( 1-\frac{2M}{r}+\frac{Q^{2}}{r^{2}}\right) dt^{2}+2\sqrt{\frac{%
2M}{r}-\frac{Q^{2}}{r^{2}}}\,dtdr+dr^{2}+r^{2}d\Omega ^{2}.  \label{rnp}
\end{equation}

Following the standard procedure, the imaginary part of the action for an
outgoing massless neutral particle can be computed, and the resulting
tunneling probability is
\begin{eqnarray}
\Gamma &\sim &\exp \left[ -4\pi \left( 2EM-E^{2}-\left( M-E\right) \sqrt{%
(M-E)^{2}-Q^{2}}+M\sqrt{M^{2}-Q^{2}}\,\right) \right]  \notag \\
&=&\exp \left( \Delta S\right) ,  \label{rntp}
\end{eqnarray}%
where $S=\pi( M+\sqrt{M^{2}-Q^{2}}\,\,)^{2}$ is the entropy. The
corresponding temperature for a charged black hole is $T=\frac{1}{2\pi }%
\frac{\sqrt{M^{2}-Q^{2}}}{( M+\sqrt{M^{2}-Q^{2}}\,\,)^{2}}$, that allows us
to compare the tunneling probability (\ref{rntp}) with the Boltzmann factor
for emission $\Gamma =\exp (-\beta E)$, $(\beta =1/T)$. As before, we find
this spectrum is non-thermal, and the relationship $\Gamma
(E_{1}+E_{2})=\Gamma (E_{1},E_{2})$ remains true. Using the same argument as
before for computing the information or entropy carried by the correlations
of two emitted particles for a Schwarzchild black hole outlined in the
previous subsection, we find
\begin{equation}
\ln \Gamma (E_{1}+E_{2})-\ln \left[ \Gamma (E_{1})\ \Gamma (E_{2})\right]
\neq 0.
\end{equation}
Again we find the existence of correlations among Hawking radiations because
the spectrum for emission from a charged black hole is also non-thermal. In
the process of two particles tunneling with respective energies $E_{1}$ and $%
E_{2}$, we find the entropy form
\begin{eqnarray}
S(E_{1}) &=&-\ln \Gamma (E_{1})=4\pi \left( 2E_{1}M-E_{1}^{2}-\left(
M-E_{1}\right) \sqrt{(M-E_{1})^{2}-Q^{2}}+M\sqrt{M^{2}-Q^{2}}\right),  \notag
\\
S(E_{2}|E_{1}) &=&-\ln \Gamma (E_{2}|E_{1})  \notag \\
&=&4\pi \left(2E_{2}\left( M-E_{1}\right) -E_{2}^{2}-\left(
M-E_{1}-E_{2}\right) \sqrt{ (M-E_{1}-E_{2})^{2}-Q^{2}}+M\sqrt{M^{2}-Q^{2}}%
\right).  \notag
\end{eqnarray}
Not surprisingly, they satisfy the definition of conditional entropy, $%
S(E_{1},E_{2})=-\ln \Gamma (E_{1}+E_{2})=S(E_{1})+S(E_{2}|E_{1})$. After a
detailed calculation, again we find that the amount of correlation is
exactly equal to the mutual information described in Eq. (\ref{tmi}), and
this shows that the correlation can carry information from a Reissner-Nordstr%
\"{o}m black hole and any single step of the emission must be entropy
preserving. We can count the total entropy carried away by all outgoing
particles, and find
\begin{equation}
S(E_{1},E_{2},\cdots,E_{n})=\sum\limits_{i=1}^{n}S(E_{i}|E_{1},E_{2},\cdots
,E_{i-1})=S(E),
\end{equation}
where $E=\sum_{i=1}^{n}E_{i}$ is the total energy of the black hole
radiation.

A subtle difference arises from the previously considered Schwarzchild black
hole. In the present case, the mass of a black hole can never decrease to
zero, yet the tunneling probability must remain a real value, thus $%
M^{2}-Q^{2}\geqslant 0$ is to be enforced. Because the tunneled particles
are taken as neutrals, the extreme case is reached when $M^{2}=Q^{2}$. The
temperature is $T=0$ in the extreme case, so Hawking radiation will vanish
and the black hole is stablized. This is consistent with cosmic censorship.
For the information loss paradox we seek to resolve, it is important to ask:
is entropy still conserved in the extreme case? The key point to a
legitimate answer depends on how to properly describe the entropy of an
extreme black hole. According to the definition of Bekenstein-Hawking, the
entropy of an extreme black hole is proportional to its surface area,
although its temperature is zero. Thus we can conclude that even in the
extreme case the entropy remains conserved. Although the entropy (or the
information ) cannot be taken out completely, the residual information
remains inside the extreme black hole. This is reasonable because the
extreme limit can be viewed as the ground state of a charged black hole,
which has a high degeneracy $\thicksim e^{S_{e}}$ with $S_{e}=\pi M^{2}$
\cite{sl05}.

A microscopic picture of tunneling by charged particles is more appropriate
in order to avoid the extreme case. Fortunately, the tunneling probability
of charged massive particles for a Reissner-Nordstr\"{o}m black hole has
been obtained in Ref. \cite{jz05}. With outgoing particles capable of
carrying away charges from a charged black hole, its semi-classical
trajectories have to be modified due to electromagnetic forces. In the
treatment of Ref. \cite{jz05}, the 4-dimensional electromagnetic potential $%
A_{\mu }=(A_{t},0,0,0)$ where $A_{t}=-Q/r$ is introduced in the quasi-Painlev%
\'{e} coordinates (\ref{rnp}) and the electromagnetic interaction $%
-(1/4)F_{\mu \nu }F^{\mu \nu }$, which can be described by the potential $%
A_{\mu }$, also has to be considered in calculating the action. Taking into
account the modifications to the equation of motion due to the change of
charge and including the contribution from the electromagnetic interaction,
the tunneling probability is found to be \cite{jz05},
\begin{eqnarray}
\Gamma &\sim &\exp \left[ \pi \left( M-E+\sqrt{(M-E)^{2}-\left( Q-q\right)
^{2}}\,\right) ^{2}-\pi \left( M+\sqrt{M^{2}-Q^{2}}\,\right) ^{2}\right]
\notag \\
&=&\exp \left( \Delta S\right) ,
\end{eqnarray}%
where $\Delta S=S(M-E,Q-q)-S(M,Q)$ is the difference of entropies for a
Reissner-Nordstr\"{o}m black hole before and after the emission, and $q$ is
the charge that is carried away by the particle with energy $E$. Comparing
with the Boltzmann factor, we find clearly this remains a non-thermal
spectrum. Using our method outlined before, we conclude there exists
information-carrying correlations among emitted particles. Analogously,
after a detailed calculation, we find that the total entropy carried away by
the outgoing particles plus that of the accompanying black hole remains
conserved, which is now due to both energy and charge conservations. Thus,
we find once again no information is lost in the Hawking radiation as
tunneling for a Reissner-Nordstr\"{o}m black hole.

\subsection{Kerr black hole}

For rotating black holes \cite{jc06}, a complication arises from the
frame-dragging effect of the coordinate system in the stationary rotating
spacetime. The matter field in the ergosphere near the horizon must be
dragged by the gravitational field with an azimuthal angular velocity. A
proper physical picture thus must be capable of describing such effects in
the dragged coordinate system. Adopting the quasi-Painlev\'{e} time
transformation and the dragging coordinate transformation for the Doran form
of the Kerr coordinates \cite{cd00}, the so-called dragged Painlev\'{e}-Kerr
coordinates can be expressed as
\begin{equation}
ds^{2}=\frac{\bigtriangleup \Sigma }{\left( r^{2}+a^{2}\right)
^{2}-\bigtriangleup a^{2}\sin ^{2}\theta }\,dt^{2}-\frac{\Sigma }{r^{2}+a^{2}%
}dr^{2}-2\frac{\sqrt{2Mr(r^{2}+a^{2})}\Sigma }{\left( r^{2}+a^{2}\right)
^{2}-\bigtriangleup a^{2}\sin ^{2}\theta }\,dtdr-\Sigma\, d\theta ^{2},
\end{equation}%
where $a$ is the angular momentum of a unit mass, $\Sigma =r^{2}+a^{2}\cos
^{2}\theta $, and $\bigtriangleup =$ $r^{2}+a^{2}-2Mr$. In this spacetime
the event horizon and the infinite red-shifted surface coincides with each
other, so that the WKB approximation can be used to calculate the imaginary
part of the action. The tunneling probability for a rotating black hole is
then found to be \cite{jc06}
\begin{eqnarray}
\Gamma &\sim &\exp \left[ -2\pi \left( M^{2}-\left( M-E\right) ^{2}+M\sqrt{%
M^{2}-a^{2}}-\left( M-E\right) \sqrt{\left( M-E\right) ^{2}-a^{2}}\,\right) %
\right]  \notag \\
&=&\exp \left( \Delta S\right) ,  \label{kbtp}
\end{eqnarray}
where $\Delta S=S(M-E,Q-q)-S(M,Q)$ is the difference of the entropies for a
Kerr black hole before and after the emission of a particle with energy $E$.
This spectrum (\ref{kbtp}) is once again non-thermal, and the interesting
property $\Gamma (E_{1}+E_{2})=\Gamma (E_{1},E_{2})$ remains true. The total
amount of correlations hidden inside Hawking radiations can again be
computed analogously, and we find
\begin{equation}
\ln \Gamma (E_{1}+E_{2})-\ln \left[\Gamma (E_{1})\ \Gamma (E_{2})\right]
\neq 0\text{.}
\end{equation}
For tunneling of two particles with respective energies $E_{1}$ and $E_{2}$,
we find the entropies
\begin{eqnarray}
S(E_{1})=-\ln \Gamma (E_{1}) &=&2\pi \left( M^{2}-\left( M-E_{1}\right)
^{2}+M\sqrt{M^{2}-a^{2}}-\left( M-E_{1}\right) \sqrt{\left( M-E_{1}\right)
^{2}-a^{2}}\,\right) , \\
S(E_{2}|E_{1})=-\ln \Gamma (E_{2}|E_{1}) &=&2\pi \left[ \left(
M-E_{1}\right) ^{2}-\left( M-E_{1}-E_{2}\right) ^{2}+\left( M-E_{1}\right)
\sqrt{\left( M-E_{1}\right) ^{2}-a^{2}}\,\right]  \notag \\
&&-2\pi \left[ \left( M-E_{1}-E_{2}\right) \sqrt{\left( M-E_{1}-E_{2}\right)
^{2}-a^{2}}\,\right] .
\end{eqnarray}%
Clearly they also satisfy the definition of conditional entropy $%
S(E_{1},E_{2})=-\ln \Gamma (E_{1}+E_{2})=S(E_{1})+S(E_{2}|E_{1})$. A
detailed calculation again reveals that the amount of correlation in this
case is exactly equal to the mutual information described in Eq. (\ref{tmi}%
). We can count the total entropy carried away by the outgoing particles in
the same manner and find
\begin{equation}
S(E_{1},E_{2},...,E_{n})=\sum\limits_{i=1}^{n}S(E_{i}|E_{1},E_{2},\cdots
,E_{i-1})=S(E)\text{,}
\end{equation}%
where $E=\sum_{i=1}^{n}E_{i}$ is the total energy of the Hawking radiations.

After a detailed calculation, we again find that no information is lost as
the tunneling process is an entropy conserving one. However, since the black
hole is rotating, angular momentum conservation must be considered. In the
tunneling process considered here, we do not see how the angular momentum is
carried away. An obvious reason is that the total angular momentum of a
black hole is absent in the coordinates %\cite{kbh}
and instead the angular momentum of a unit mass is used. The outgoing
particles clearly carry away the angular momentum and this can be seen in
the calculation for the imaginary part of the action and the tunneling
probability. We conclude that the entropy conservation is due to both energy
conservation and angular momentum conservation. Once again, it is the
existence of information-carrying correlations due to energy conservation
that resolves the information loss paradox for Hawking radiation as
tunneling for a Kerr black hole.

We note that due to the angular momentum conservation of a unit mass in the
tunneling process, the extreme case $a^{2}=M^{2}$ is guaranteed to appear
and the radiation will then stop. However this does not contradict our
conclusion because in the extreme limit all tunneling processes vanish and
the residual entropy will remain inside the extreme black hole, which is
consistent with entropy conservation and unitarity.

\subsection{Kerr-Newman black hole}

If we were to consider uncharged massless particles tunneling from a charged
rotating black hole, the calculation would completely parallel that for a
Kerr black hole. Instead, we consider tunneling of charged massive particles
for a Kerr-Newman black hole \cite{jc06,jz06}. Like for a Kerr black hole,
one has to treat the frame-dragging effect using dragging coordinate
transformation, and then the event horizon becomes consistent with the
infinite red-shifted surface, which allows for the WKB approximation to be
used. On the other hand, because the particles are now charged, the
contribution to the action from the electromagnetic interaction has to be
included.

Performing the quasi-Painlev\'{e} time transformation and the dragging
coordinate transformation for the Boyer-Lindquist form of the Kerr-Newman
coordinates \cite{ncc65}, the dragged Painlev\'{e}-Kerr-Newman coordinates
are obtained as following

\begin{equation}
ds^{2}=\frac{\bigtriangleup \Sigma }{\left( r^{2}+a^{2}\right)
^{2}-\bigtriangleup a^{2}\sin ^{2}\theta }\,dt^{2}-\frac{\Sigma }{r^{2}+a^{2}%
}\,dr^{2}-2\frac{\sqrt{\left( 2Mr-Q^{2}\right) (r^{2}+a^{2})}\,\Sigma }{%
\left( r^{2}+a^{2}\right) ^{2}-\bigtriangleup a^{2}\sin ^{2}\theta }%
\,dtdr-\Sigma\, d\theta^{2}\text{.}
\end{equation}
Calculating the imaginary part of the action in the usual manner including
the contribution from electromagnetic interaction, the tunneling probability
of charged massive particles for a Kerr-Newman black hole \cite{jc06,jz06}
is found to be
\begin{eqnarray}
\Gamma &\sim &\exp \left[ \pi \left( M-E+\sqrt{(M-E)^{2}-\left( Q-q\right)
^{2}-a^{2}}\,\right)^{2}-\pi \left( M+\sqrt{M^{2}-Q^{2}-a^{2}}\,\right) ^{2}%
\right]  \notag \\
&=&\exp \left( \Delta S\right) \text{,}  \label{knbtp}
\end{eqnarray}
where $\Delta S=S(M-E,Q-q)-S(M,Q)$ is the difference of entropies for a
Kerr-Newman black hole before and after the emission of a particle with
energy $E$ and charge $q$. Not surprisingly, the spectrum is again a
non-thermal one, and an analogous relationship $\Gamma
(E_{1}+E_{2},q_{1}+q_{2})=\Gamma (E_{1},q_{1},E_{2},q_{2})$, from which we
can affirm the existence of correlation between two radiated emissions as
\begin{equation}
\ln \Gamma (E_{1}+E_{2},q_{1}+q_{2})-\ln \left[\Gamma (E_{1},q_{1})\ \Gamma
(E_{2},q_{2})\right] \neq 0\text{.}
\end{equation}%
Like other types of black holes we considered previously, we find that there
exists correlation in Hawking radiation from a charged rotating black hole.
For two emissions with respective energies, $E_{1}$ and $E_{2}$, and
charges, $q_{1}$ and $q_{2}$, we find
\begin{eqnarray}
S(E_{1},q_{1})=-\ln \Gamma (E_{1},q_{1}) &=& \pi \left( M-E_{1}+\sqrt{%
(M-E_{1})^{2}-\left( Q-q_{1}\right) ^{2}-a^{2}}\,\right) ^{2}  \notag \\
&&-\pi \left( M+\sqrt{M^{2}-Q^{2}-a^{2}}\,\right) ^{2},
\end{eqnarray}
\begin{eqnarray}
S(E_{2},q_{2}|E_{1},q_{1})=-\ln \Gamma (E_{2},q_{2}|E_{1},q_{1}) &=&\pi
\left( M-E_{1}-E_{2}+\sqrt{(M-E_{1}-E_{2})^{2}-\left( Q-q_{1}-q_{2}\right)
^{2}-a^{2}}\,\right) ^{2}  \notag \\
&&-\pi \left( M-E_{1}+\sqrt{\left( M-E_{1}\right) ^{2}-Q^{2}-a^{2}}\,\right)
^{2}.
\end{eqnarray}
They satisfy the definition for conditional entropy $%
S(E_{1},q_{1},E_{2},q_{2})=-\ln \Gamma
(E_{1}+E_{2},q_{1}+q_{2})=S(E_{1},q_{1})+S(E_{2},q_{2}|E_{1},q_{1})$. After
a detailed calculation, we again find the amount of correlation is exactly
equal to the mutual information described in Eq. (\ref{tmi}). We can count
the total entropy carried away by the outgoing particles in the same manner,
and we find
\begin{equation}
S(E_{1},q_{1},E_{2},q_{2},...,E_{n},q_{n})=\sum%
\limits_{i=1}^{n}S(E_{i},q_{i}|E_{1},q_{1},E_{2},q_{2},\cdots
,E_{i-1},q_{i-1})=S(E,q)\text{,}
\end{equation}%
where $E=\sum_{i=1}^{n}E_{i}$ and $q=\sum_{i=1}^{n}q_{i}$ is the total
energy and charge of the Hawking radiations. This shows entropy conservation
in the Hawking radiation for a Kerr-Newman black hole. Clearly the
conservation of entropy arises because of energy conservation, charge
conservation, and angular momentum conservation being rightfully enforced
for the process. We thus find no information is lost in Hawking radiation as
tunneling for a Kerr-Newman black hole. Like the situation for the Kerr
black hole, the angular momentum of a unit mass remains a constant, or
conserved in the tunneling process, the extreme case $a^{2}=M^{2}$ thus must
appear in the end of tunneling process. However, this doesn't contradict our
conclusions because the Hawking radiation terminates in the extreme case,
and the residual entropy will remain a constant inside the black hole.

Before concluding this section, we will explain why information can be
carried away from a black hole by Hawking radiation as tunneling. The most
important reason is that the emission process is probabilistic, not a
deterministic one. For each tunneling emission from a black hole, we only
know a radiation may occur with a probability $\Gamma (E)$, nothing else. In
other words, the uncertainty of the event (for a radiation with energy $E$)
or the potential information we can gain from the event is $S(E)=-\ln \Gamma
(E) $.

When a radiation with energy $E_{1}$ is received, the potential information
we can gain is $S(E_{1})=-\ln \Gamma (E_{1})$. After already receiving the
emission at energy $E_{1}$, when we receive the next radiation with energy $%
E_{2}$, the potential information we can gain is $S(E_{2}|E_{1})=-\ln \Gamma
(E_{2}|E_{1})$. Step by step, we can track each subsequent emission, all of
the $n$ emissions until the black hole stops radiating. Of course, we have
to assume that the observer is rightfully equipped to detect all radiations.
In the end, the information gained by the observer is $%
S(E_{1},E_{2},...,E_{n})$, where for a Schwarzchild black hole, $%
S(E_{1},E_{2},...,E_{n})$ is nothing but the initial entropy of the black
hole. For a Reissner-Nordstr\"{o}m black hole, a Kerr black hole, and a
Kerr-Newman black hole, the entropy of their extreme black holes is $%
S_{e}=S_{\mathrm{BH}}-S(E_{1},E_{2},...,E_{n})$, where
$S_{\mathrm{BH}}$ is their respective initial entropy. However,
given all radiations are received and detected by an observer,
he/she cannot reconstruct the initial state from which the matter
collapses into a black hole. A legitimate reconstruction may need
more knowledge concerning the dynamic description for black hole
radiation, which seems only possible with a complete theory for
quantum gravity.

In concluding this section, we use statistical method and quantum
information theory to show that no information is lost in Hawking radiation
as tunneling. This result constrains the black hole evaporation as tunneling
to be a unitary process. Our conclusion is based on a single but an
important observation, that a non-thermal Hawking radiation spectrum implies
the existence of information-carrying correlation among emitted particles.
Upon counting the total entropy, the portion due to correlation is found to
exactly balance the part previously perceived as lost. Within our suggested
resolution, we find that energy conservation or the self-gravitation effect
plays a crucial role. For some black holes such as charged, rotating, and
charged-rotating black holes, other conservation laws, such as charge
conservation and angular momentum conservation, also affect the tunneling
rate and are important for the entropy conservation. This implies that in
further studies about the black hole information loss paradox, quantum
gravity theory should be considered within the framework of energy, charge,
and angular momentum conservations.

\section{Hawking radiation as \textbf{tunneling through a quantum horizon}}

Hawking radiation as tunneling through a quantum horizon has been considered
before \cite{amv05}, and the tunneling probability is already given in a
general spherically symmetric system in the ADM form \cite{kw95} by
referencing to the first law of black hole thermodynamics $dM=\frac{\kappa }{%
2\pi }dS$,
\begin{equation}
\Gamma \sim (1-\frac{E}{M})^{2\alpha }\exp \left[ -8\pi E\left( M-\frac{E}{2}%
\right) \right] =\exp \left( \Delta S\right) ,  \label{qtpo}
\end{equation}%
where $S=\frac{A}{4}+\alpha \ln A$ is the entropy derived by directly
counting the number of micro-states with string theory and loop quantum
gravity \cite{amv05}. The coefficient $\alpha $ is negative in loop quantum
gravity \cite{gm05}. Its sign remains uncertain in string theory, depending
on the number of field species in the low energy approximation \cite{sns98}.
When $\alpha >0$ and the tunneling energy approaches the mass of the black
hole, the tunneling probability $\Gamma \rightarrow 0$. A more interesting
case occurs when $\alpha <0$, the black hole will not radiate away all of
its mass. The tunneling will halt at a critical value of the black hole mass
giving rise to a situation similar to a black hole remnant as described in
Ref. \cite{lx07}. Recently, some quantum properties of the
Bekenstein-Hawking entropy and its universal sub-leading corrections have
been discussed \cite{sh04,ajmm04,cpb07}.

For Hawking radiation as tunneling through a quantum horizon as described by
the Eq. (\ref{qtpo}), we can use the same statistical method as described by
Eqs. (\ref{1tp}) and (\ref{atp}) to probe and measure correlation. We find
as before the interesting relationship $\Gamma (E_{1}+E_{2})=\Gamma
(E_{1},E_{2})$ remains true. The amount of correlation is evaluated to be
\begin{equation}
\ln \Gamma (E_{1}+E_{2})-\ln \left[ \Gamma (E_{1})\ \Gamma (E_{2})\right]
=8\pi E_{1}E_{2}+2\alpha \ln \frac{M(M-E_{1}-E_{2})}{(M-E_{1})(M-E_{2})}\neq
0.  \label{qco}
\end{equation}%
Once again, correlations are found to exist due to the non-thermal nature of
the spectrum Eq. (\ref{qtpo}), despite being corrected by quantum gravity
effect. Unlike the result of Eq. (\ref{co}), an additional term appears as
the second term before the last inequality sign in Eq. (\ref{qco}). On
careful examination, we conjecture that this correction may carry
information about effects of quantum gravity or black hole area quantization
\cite{sh04,ajmm04,cpb07}.

The process of two emissions can be considered as in the situation when the
Bekenstein-Hawking entropy is used, and we find
\begin{eqnarray}
S(E_{1})=-\ln \Gamma (E_{1}) &=&8\pi E_{1}(M-\frac{E_{1}}{2})-2\alpha \ln (1-%
\frac{E_{1}}{M}), \\
S(E_{2}|E_{1})=-\ln \Gamma (E_{2}|E_{1}) &=&8\pi E_{2}(M-E_{1}-\frac{E_{2}}{2%
})-2\alpha \ln \left( 1-\frac{E_{2}}{M-E_{1}}\right) .
\end{eqnarray}%
Again this form is consistent with the definition of conditional entropy $%
S(E_{1},E_{2})=S(E_{1})+S(E_{2}|E_{1})$. Repeating the steps until the black
hole is exhausted by emissions, we find
\begin{equation}
S(E)=\sum\limits_{i=1}^{n}S(E_{i}|E_{1},E_{2},\cdots ,E_{i-1}),  \label{qtp}
\end{equation}%
where $E=\sum_{i=1}^{n}E_{i}$ is the total energy of the black hole
radiation.

For $\alpha >0$, we find $\Gamma (E)\to 0$ when $E\to M$, but $S\left(
M-E\right)\to\infty$. This causes difficulty explaining the origin of an
exponentially growing entropy when the black hole vanishes. However,
qualitatively, this actually can be understood within the picture of Hawking
radiation from a black hole. In the limit of $\Gamma (M)$ $=0$, the
tunneling energy approaches the mass of the black hole, and the tunneling
becomes slower and slower while the time to exhaust a black-hole approaches
infinite. This infinity also can be obtained from other methods by using the
Stefan-Boltzmann law as in Ref. \cite{fp07}.

For $\alpha <0$, it is known \cite{lx07} that when the mass of a black hole
approaches the critical mass $M_{c}$, no particles will be emitted. From Eq.
(\ref{qtp}) we then obtain $S(M)-S(M_{c})=\sum%
\limits_{i}S(E_{i}|E_{1},E_{2},\cdots ,E_{i-1})$ or $S(M)=\sum%
\limits_{i}S(E_{i}|E_{1},E_{2},\cdots ,E_{i-1})+S(M_{c})$. In Ref. \cite%
{lx07}, the mass $M_{c}$ is called the \textquotedblleft zero point
energy\textquotedblright\ of a black hole that is similar to a black hole
remnant because it does not depend on the initial black hole mass. We have
shown that even with such a remnant, the total entropy remains conserved
when information carried away by correlations are correctly included. Thus
the unitarity remains true when the classical horizon is replaced by a
quantum one for Hawking radiation.

\section{\textbf{Conclusions}}

In this work, we have significantly expanded our self-consistent theory for
the resolution of the paradox for black hole information loss. In the
picture of Hawking radiation as tunneling \cite{pw00}, we have earlier
pointed out the existence of correlations among radiations whenever the
emission spectrum or the tunneling probabilities are non-thermal. While
phenomenological, this resolution, first proposed by us in an earlier
communication \cite{zcz09}, is firmly supported by statistical theory.
Although we reply on the results from a semi-classical treatment of Hawking
radiation as tunneling, there is no room for comprise regarding our main
conclusion for the existence of correlation as we have shown here for the
various type of black holes. Whenever the Hawking radiation spectrum takes a
non-thermal form, correlations must exist among radiated particles,
irrespective of the nature for these emissions being charged or neutral,
massive or massless particles, etc.

By comparing the amount of information that can be encoded into this
correlation with the mutual information of quantum information
theory as we did previously \cite{zcz09}, we have shown in this work
for an extensive list of black holes that the amount of information
is always precisely equal to the mutual information. Due to the
tunneling or the emission, the mass of a black hole decreases, that
lowers the entropy of a black hole. According to the general second
law of thermodynamics, the total entropy of the system, consisting
of a black hole and its radiation, can never decrease. Thus, the
tunneled particles as Hawking radiation must carry away entropy.
Upon careful evaluation of the total entropy for radiated particles,
including the contribution of the correlation, we find that the
total entropy is conserved in the tunneling process, which supports
the statement of unitary evolution for Hawking radiation as
tunneling \cite{zcz09}. In addition to the standard Schwarzschild
black holes we considered earlier \cite{zcz09}, the list of black
holes we consider in this article includes Reissner-Nordstr\"{o}m
black holes, Kerr black holes, and Kerr-Newman black holes.
Surprisingly, or perhaps not so surprisingly, even when considering
the corrections due to quantum gravity \cite{zheda} in the tunneling
process, our conclusions remain true: there exists correlation among
Hawking radiations if the emission spectrum is non-thermal; the
total amount of correlation exactly balances the mutual entropy; the
Hawking radiation is an entropy conserving process. Within the
framework of our result, we shall view the tunneling particles and
the correlations as messengers capable of carrying away entropy and
information to assure that the entropy for the total system is
conserved.

Based on our current understanding, we feel our conclusions are significant
at least in two aspects. The first concerns black hole thermodynamics. It
has been shown before that the tunneling process we discuss satisfies the
first law of black hole thermodynamics irrespective of whether the horizon
is classical \cite{tp08} or quantum \cite{zcz08}. No conclusive consensus
exist concerning the second law of black hole thermodynamics. Before our
work, it was not known how entropy changes, and our conclusion that Hawking
radiation is an entropy conserving process is thus quite exciting. This
opens the door to reversibility and unitary dynamics for a black hole in
principle. We note that reversibility is consistent with microscopic
unitarity. The second aspect concerns the information loss paradox. There
are two important results here: (1) we have shown within the picture of
Hawking radiation as tunneling that the entropy growth can be exactly
balanced, which provides a necessary condition to resolve the Hawking
paradox of black hole information loss. Any legitimate attempt to resolve
this paradox must find a solution to balance the entropy growth during the
black hole evaporation process if individual emissions are considered
independent. If entropy is indeed growing in a process, this process cannot
be unitary. (2) We have shown that there exists correlation among Hawking
radiations from a black hole. Using quantum information theory, the mutual
information between two quantum subsystems or two emitted particles as
considered in this study, we show that the amount of correlation among the
black hole radiations can encode exactly the same amount of information and
carry them away upon emission, which certainly represents a plausible
resolution to release the information locked in a black hole. Any resolution
to the information loss paradox of a black hole requires such a suitable
mechanism to release the information locked in a black hole. Fortunately, we
have uncovered such a mechanism inherent in the correlations of emitted
particles.

In summary, we have shown that entropy is conserved in Hawking
radiation. This resolves the paradox of black hole information loss.
The amount of information that formerly was perceived to be lost is
found to be encoded and carried away by Hawking radiations. This
lends strong support to the belief that Hawking radiation as
tunneling is unitary, in principle, based on the work we presented
in this article.

As a final remark we note that for some cases we study, a black hole
does not radiate away all of its mass and leaves behind a remnant in
the end or forms an extreme black hole. Even for these special
cases, our study shows that our conclusions remain valid.
Information is still leaked out so long as a black hole starts to
radiate. Based on our theoretical treatment, or the resolution we
present for the paradox of black hole information loss due to
Hawking radiation, we find that energy conservation or
self-gravitation effect should be enforced when quantum theory is
unified with gravity. Conservation laws are the most fundamental
elements in a consistent theory for quantum gravity.

\section{\textbf{Acknowledgement}}

This work is supported by NSFC under Grant No. 11074283.

\end{document}